# Chandra Publication Statistics

*Arnold H. Rots, Sherry L. Winkelman, Glenn E. Becker*
*CXC / Smithsonian Astrophysical Observatory*

**Abstract**
In this study we develop and propose publication metrics, based on an analysis of data from the Chandra bibliographic database, that are more meaningful and less sensitive to observatory-specific characteristics than the traditional metrics. They fall in three main categories: speed of publication; fraction of observing time published; and archival usage. Citation of results is a fourth category, but lends itself less well to definite statements. For Chandra, the median time from observation to publication is 2.36 years; after about 7 years 90% of the observing time is published; after 10 years 70% of the observing time is published more than twice; and the total annual publication output of the mission is 60-70% of the cumulative observing time available, assuming a two year lag between data retrieval and publication.

## 1. Introduction: Scope and Objectives

The Chandra X-ray Observatory[1] ("Chandra"; Weisskopf et al. 2002) was launched in July 1999 as the third of NASA's Great Observatories. It provides high-resolution (< 1 arcsecond) imaging, and moderately high-resolution (Q up to 1000) spectroscopy in the energy range 0.2 to 10 keV. The amount of science exposure time available on the observatory is typically 20 Ms per year, yielding a science observing efficiency of about 65%. Table 1 provides, for reference and context, the distribution of exposure time over scientific research areas of interest. Early in the mission it was decided to maintain in the Chandra Data Archive[2] an extensive bibliographic database[3] for this mission, holding a large variety of metadata on all papers related to Chandra and, importantly, explicit links between datasets in the archive and the publications presenting those data. For a comparison of the practices concerning bibliographic databases at 14 missions and observatories, see Lagerstrom (2010).

In this paper we take a careful look at the information in Chandra's bibliographic database to discern the habits of Chandra observers and the observatory's achievements in engendering the production of important scientific publications. One should bear in mind that bibliographic databases will never be perfect. The Chandra bibliographic database is created, populated, and maintained by the Chandra Archive Operations Team using a mix of automated, manual, and visual labor – all three in generous amounts. We are confident that the database has attained an acceptable level of reliability and we will continue to improve its accuracy; but it means that the numbers are never truly definitive and are likely to improve as time goes on.

---

[1] http://cxc.harvard.edu/
 http://cxc.harvard.edu/biblio/chandra_bib.html
[2] http://cxc.harvard.edu/cda/
[3] http://cxc.harvard.edu/cgi-gen/cda/bibliography

Table 1. Use of Chandra Exposure Time by Research Area for Proposal Cycles 1 through 11

| Research Area | Percentage Exposure Time |
|---|---|
| Solar System | 0.7 |
| Stars and White Dwarfs | 14.2 |
| White Dwarf Binaries and Cataclysmic Variables | 3.9 |
| Black Holes and Neutron Star Binaries | 7.4 |
| Supernovae, Supernova Remnants, and Isolated Neutron Stars | 15.9 |
| Galactic Diffuse Emission and Surveys | 2.8 |
| Normal Galaxies | 10.1 |
| Active Galaxies and Quasars | 19.6 |
| Clusters of Galaxies | 16.3 |
| Extragalactic Diffuse Emission and Surveys | 9.1 |

In order to ensure as high a level of reliability as possible, we will first address the question whether there are publication metrics that are more suitable than the citation counts which are commonly the basis for the metrics used in these kinds of studies (see, e.g., Grothkopf & Lagerstrom 2011; Apai et al. 2010; Trimble & Ceja 2007, 2008, and 2010; Trimble 2009; Madrid & Macchetto 2009; Crabtree 2008; Abt 2003; Benn 2002). We will conclude with recommending some metrics that can be used across multiple observatories in comparative studies as a more informative substitute for the citation counts or simple paper counts.

To help frame the issue of cross-observatory comparisons, it may be good to point out some of the pitfalls. Aside from the question whether the metrics, such as number of papers and number of citations, are derived on the basis of criteria that are applied homogeneously across the observatories, there are properties intrinsic to the observatories themselves that have significant impact on these numbers and, thereby, limit the accuracy and usefulness of the comparisons. The actual statistics are affected by factors such as:

- The age of the observatory and the longevity of its publication list
- The size of the observatory's constituency and the amount of research funding it receives
- The breadth of research projects for which the observatory is relevant
- The number of observations that can be made in a year; folded into this are parameters such as scheduling constraints, calibration requirements, bandwidth, sensitivity, effective aperture, and typical source flux density
- The uniqueness of the data and their value for archival research

The temptation to use favorable publication statistics in defense of one's observatory is understandable, but in making comparisons across observatories one should be cognizant of such effects. One of our objectives in this paper is to establish metrics that are not (or are, at least, less) sensitive to these factors.

## 2. Choice of Metrics

Measuring the success or impact of an observatory through bibliographic data is a hazardous undertaking, not only because of the difficulty of defining meaningful metrics, but also because of the impossibility to unambiguously define key concepts in a way that has validity across observatories. See also Grothkopf & Lagerstrom (2011), Abt (2003), and Benn (2002) for a discussion of these issues. We will highlight some of the most essential concepts and make recommendations.

### 2.1. Observations

The concept of an "observation" not only varies widely between observatories, but is even difficult to define in the context of a single mission, as will become clear in our discussion of data selection. Our recommendation is to adopt a consistent definition that is *reasonable* within the context of the observatory and to only use it sparingly for derived parameters, such as the median publication delays (cf. Figs. 1 and 3), that are not very sensitive to the exact definition.

Rather than using the number of observations, and what fraction of them is covered in publications, as a measure of an observatory's output, we strongly recommend the use of exposure time, as it is a more informative measure of effectiveness and efficiency – or, if one prefers, good stewardship of available resources. This should be restricted to science exposure time, excluding exposure time spent on engineering measurements and calibration observations. Note that these statistics, particularly fraction of available exposure time, can be refined by calculating them, for instance, by instrument.

Along similar lines, a simple count such as the number of refereed papers published per year is not a good metric. We suspect that early in a mission, each observation tends to result in a paper, while later on more and more observations are combined into single papers. For Chandra this is corroborated by the fact that the number of refereed scientific papers has remained fairly constant after the first three years of the mission, but that the amount of exposure time published in those papers has continued to increase, year after year. During 2001 and 2002 the percentage of single-observation papers was 58% and average number of observations per paper 2.9; by 2008 and 2009 single-observation papers constituted 36% and the average number of observations had increased to 11.5. We have no reason to believe that it is any different for other missions and therefore recommend using "exposure time published" as the more meaningful metric.

The definition of a paper "that presents an observation" is crucial for this metric to work. This is, admittedly, a somewhat grey area. The two criteria we have used are that the paper must provide an unambiguous link to a specific observation and that some quantity or property was derived from that observation (rather than, for instance, just quoting the result from a previous paper). Although the AAS journals allow (and encourage) authors to insert the links to the data into their manuscript through the use of Dataset Identifiers, very few authors take advantage of this mechanism and virtually all links for Chandra papers are created by the Archive Operations Team.

### 2.2. Journals

It goes without saying that only refereed papers should be counted for any metric. The Chandra bibliographic database covers all publications that are indexed by the Astrophysics Data System[4] (ADS) and accepts the ADS's designation of refereed publications. One may argue whether the ADS made the right choice in all cases, but at least this is an unambiguous criterion that can be shared among missions.

The net should be cast wide when it comes to journals covered and inter-observatory comparisons should not be restricted to a subset of journals since different communities tend to have different habits in this respect. We cover all journals that are so designated by the ADS, with the exception of review article journals, conference proceedings journals, and observatory publications. Table 2 provides all refereed journals used in this study, with the percentage of science papers for each of them for Chandra and 12 other observatories. Table 3 provides the translation of observatory keys (in Table 2) to observatory names, the period covered for each, and some summary information.

---

[4] http://adsabs.harvard.edu/

We have split the journals in three categories: basic core journals (the ones everyone thinks of as the prime professional journals); core–90 journals (the set of journals that have published at least 90% of all papers for all observatories); and the remaining journals that contain Chandra publications. The table clearly shows how different communities gravitate toward different journals beyond the basic core set. The results for the basic core journals indicate that the publications of very high energy observatories are underrepresented in the astronomical journals.  When one expands the list to cover the core–90 journals, they fare better, but there remains a significantly significant spread between observatories in the percentage of papers covered.

Table 2. Distribution of Papers for 13 Observatories over Journals (%)

| Journal | Observatory Key (see Table 3) | | | | | | | | | | | | |
|---|---|---|---|---|---|---|---|---|---|---|---|---|---|
| | A | B | C | D | E | F | G | H | I | J | K | L | M |
| ApJ | 53.9 | 45.3 | 51.1 | 33.0 | 36.6 | 39.7 | 33.9 | 34.2 | 56.1 | 46.2 | 50.1 | 20.1 | 41.5 |
| ApJS | 2.3 | 3.3 | 5.4 | 1.2 | 0.7 | 1.3 | | 1.7 | 0.5 | 2.0 | 3.4 | 1.4 | 3.2 |
| A&A | 17.2 | 15.3 | 18.0 | 35.1 | 12.6 | 23.6 | 7.8 | 29.9 | 17.5 | 10.1 | 16.9 | 52.9 | 19.0 |
| A&AS | | 0.8 | | 0.0 | | | | 2.1 | | 0.4 | 11.6 | | |
| AJ | 3.8 | 10.0 | 6.8 | 1.9 | 0.2 | 1.5 | 0.2 | 4.9 | 0.3 | 1.4 | 0.8 | 3.9 | 9.2 |
| MNRAS | 14.2 | 15.7 | 13.0 | 20.2 | 12.7 | 23.3 | 11.8 | 17.0 | 20.2 | 13.2 | 4.8 | 17.0 | 15.5 |
| PASP | 0.2 | 1.4 | 1.1 | 0.1 | | 0.5 | | 1.3 | 0.3 | 0.3 | | 0.5 | 0.6 |
| PASJ | 1.6 | 0.5 | 0.6 | 1.3 | 0.5 | 1.1 | 39.4 | 1.4 | 0.4 | 13.9 | 0.6 | 0.0 | 0.4 |
| Nature | 0.5 | 1.1 | 0.9 | 0.4 | 0.1 | 2.0 | 0.2 | 0.5 | 0.1 | 0.2 | 0.1 | 1.1 | 0.9 |
| **Total Basic Core** | **93.7** | **93.4** | **96.9** | **93.2** | **63.4** | **93.0** | **93.3** | **93.0** | **95.4** | **87.7** | **88.3** | **96.9** | **90.3** |
| AdSpR | 1.4 | 0.2 | 0.3 | 2.0 | 0.8 | 1.2 | 1.8 | 1.7 | 1.2 | 5.4 | 3.4 | 0.0 | 0.1 |
| AN | 0.9 | 0.4 | 0.3 | 2.6 | 0.1 | 0.1 | 1.6 | 1.5 | 1.0 | 3.9 | | 0.2 | 0.8 | 1.3 |
| PhysRevD | 0.2 | 0.1 | | 0.1 | 11.3 | 0.5 | | | | | 0.3 | 0.0 | 0.3 |
| JCAP | 0.1 | 0.0 | 0.1 | 0.0 | 6.3 | 0.1 | 0.2 | 0.0 | | | | | 0.2 |
| NIMPA | | | | | 3.0 | | 0.8 | | | | | | 0.0 |
| AppPhys | 0.0 | 0.0 | | 2.0 | 0.1 | 0.2 | 0.0 | 0.0 | 0.3 | 0.9 | | 0.0 |
| Ap&SS | 0.6 | | 0.2 | 0.8 | 1.9 | 0.2 | | 0.9 | 0.5 | 0.3 | 0.6 | 0.1 | 0.9 |
| PhysRevL | 0.0 | | | | 1.8 | 0.1 | | | | | 0.6 | 0.0 | 0.1 |
| Science | 0.4 | 0.5 | 0.3 | 0.3 | 1.2 | 0.4 | 0.2 | 0.3 | | 0.1 | 0.1 | 0.3 | 0.4 |

| Journal | Observatory Key (see Table 3) | | | | | | | | | | | | |
|---|---|---|---|---|---|---|---|---|---|---|---|---|---|
| | **A** | **B** | **C** | **D** | **E** | **F** | **G** | **H** | **I** | **J** | **K** | **L** | **M** |
| RAA | 0.1 | 0.1 | 0.1 | 0.1 | 1.1 | | 0.2 | | 0.1 | | 0.1 | | 0.2 |
| Icarus | 0.1 | 2.5 | 0.7 | 0.0 | | 0.5 | | 0.1 | | | | 1.1 | 0.4 |
| **Total Core-90** | **97.5** | **97.2** | **98.9** | **99.1** | **92.9** | **96.2** | **98.3** | **97.5** | **98.2** | **97.7** | **94.5** | **99.2** | **94.2** |
| ChJAS | 0.3 | | 0.1 | 0.0 | 0.3 | 0.4 | 0.2 | | | 0.1 | | | |
| NewA | 0.3 | 0.1 | 0.1 | 0.2 | 0.2 | 0.1 | | 0.3 | | 0.3 | 0.4 | 0.3 | 0.1 |
| ChJAA | 0.2 | 0.1 | 0.1 | 0.1 | 0.2 | 0.7 | | 0.3 | 0.1 | 0.1 | 0.1 | | 0.2 |
| JGRA | 0.2 | 0.6 | | 0.1 | 0.3 | | | 0.0 | | | 0.0 | 0.0 | 0.0 |
| IJMPD | 0.2 | 0.1 | 0.1 | | 0.2 | | 0.2 | | | | | | 0.3 |
| AstL | 0.2 | 0.6 | 0.0 | 0.1 | 0.1 | | | 0.0 | 0.4 | 0.3 | 0.8 | | 0.2 |
| PhR | 0.1 | | | | 0.1 | | | | | | 0.1 | | |
| JKAS | 0.1 | 0.1 | 0.1 | | 0.1 | | | 0.1 | | 0.1 | | 0.0 | 0.1 |
| P&SS | 0.1 | 0.1 | 0.1 | 0.1 | | | | 0.1 | | | | | 0.0 |
| PhPl | 0.1 | | | | | | | | | | | | |
| GeoRL | 0.1 | 0.3 | | 0.1 | 0.4 | | | | | | 0.1 | | 0.0 |
| BASI | 0.0 | 0.0 | | | | | 0.0 | 0.0 | | | 0.3 | | 0.2 |
| PhLB | 0.0 | | | | 1.1 | | | | | | | | |
| ExA | 0.0 | | 0.1 | 0.0 | 0.1 | | | 0.0 | | | | | 0.0 |
| PASA | 0.0 | | 0.1 | 0.0 | 0.1 | 0.1 | | 0.1 | 0.1 | | 0.3 | | 0.2 |
| JSARA | 0.0 | | 0.0 | | | | | | | | | | 0.0 |
| RvMP | 0.0 | | | | | | | | | | | | 0.0 |
| AcPPB | 0.0 | | | | | | | | | | | | |
| AcA | 0.0 | 0.1 | | 0.0 | | | | | 0.3 | | 0.1 | 0.0 | 0.0 |
| JASTP | 0.0 | | | | | | | | | | | | |
| JQSRT | 0.0 | | | | | | | | | | | | |
| NCimB | 0.0 | | 0.1 | | 0.3 | 0.7 | 0.8 | | | | | | |
| NuPhA | 0.0 | | | | | | | | | | 0.1 | | |
| M&PS | 0.0 | | | | | | | | | | | | |
| Ap | 0.0 | 0.1 | 0.1 | 0.0 | | | | 0.1 | | | | 0.0 | 0.1 |
| ScChG | 0.0 | | 0.0 | | | 0.4 | | | | | | | 0.1 |
| JATP | 0.0 | | | | | | | | | | | | |
| JApA | 0.0 | | | | 0.8 | | | 0.0 | 0.2 | | 0.1 | 0.0 | 0.3 |
| NCimC | 0.0 | | | | 0.4 | | | 0.0 | | | | | |
| JAVSO | 0.0 | | | | | 0.1 | | | | | | | |
| TJPh | 0.0 | | | | | | | | | | | | |
| JPhG | 0.0 | | | | | | | | | | 0.1 | | |
| **Total** | **100.0** | **99.4** | **99.9** | **99.8** | **97.5** | **98.8** | **99.5** | **98.5** | **99.3** | **98.6** | **97** | **99.5** | **96.0** |

Table 3. Observatory Summary and Keys to Table 2

| Key | Observatory | Number of Journals | Number of Papers | Basic Core | Core-90 | Coverage |
|---|---|---|---|---|---|---|
| A | Chandra | 52 | 4564 | 93% | 97% | 2001-2011 |
| B | HST | 42 | 4924 | 92% | 95% | 2005-2011 |
| C | Spitzer | 41 | 3896 | 96% | 97% | 2001-2012 |
| D | XMM | 33 | 3629 | 93% | 98% | 2000-2011 |
| E | Fermi | 45 | 1138 | 63% | 91% | 2005-2012 |
| F | Swift | 29 | 820 | 92% | 95% | 2005-2012 |
| G | Suzaku | 20 | 501 | 93% | 98% | 2006-2012 |
| H | Rosat | 40 | 2876 | 92% | 95% | 1990-2012 |
| I | RXTE | 23 | 2427 | 95% | 97% | 1996-2012 |
| J | ASCA | 29 | 1164 | 88% | 98% | 1994-2011 |
| K | CGRO | 40 | 970 | 88% | 94% | 1992-2008,2010-2011 |
| L | ESO | 26 | 4444 | 96% | 99% | 2005-2011 |
| M | NRAO | 66 | 2334 | 89% | 93% | 2007-2011 |

### 2.3. Citations

Citation statistics are interesting and informative, but their use in a metric is treacherous for three reasons: normalization; weight; and self-citation. Again, different communities have different habits and some journals are more popular and/or more prestigious than others; that makes it hard to design a normalization algorithm that allows us to compare citation rates between observatories. The reasons for individual citations vary widely and their weight cannot easily be discerned: a paper may be cited because it has provided information that is crucial for the citing paper, but it is equally likely that a citation is rather casual. Self-citation can be detected, but it is not a binary question: if a paper with 14 authors is cited by another paper with 14 authors and they have one author in common, as well as an author with the name "J. Smith", does that count as a self-citation? Taken together, these factors make citation statistics a crude and dubious metric. Benn (2002) provides a thorough, comprehensive discussion of the disadvantages associated with citation counts, but uses them (with caution!) for lack of a better metric. See also the discussion by Grothkopf & Lagerstrom (2011).

Crabtree (2008) uses an Impact Distribution Function and in a later private communication tried to design an objective "performance factor" based on that function. Aside from the fact that it suffers from the three defects we just mentioned, it is severely flawed in its definition. This performance factor is the ratio of the number of high-citation-rate papers to low-citation-rate papers. That means that an observatory with 10 papers in both groups rates highly, while an observatory with 100 highly cited papers and 500 papers with a low citation count fares very poorly. It is not obvious to us that the former observatory performs significantly better.

As traditional citation statistics are not the most reliable metric for measuring the impact of an observatory, we suggest as a more meaningful metric of impact the number of refereed journal articles that cite, or refer to, the observatory's observations or the results of those observations; we exclude from this number the ones that also present any of its observations. In our case, we do include such papers in our bibliography, requiring that the reference is substantive, not merely a vague mentioning of Chandra's name; i.e., some fact, conclusion, or parameter that was derived from one or more Chandra observations and that has significance for the scientific contents of the article. We hasten to admit that even so, this is a somewhat dubious metric for comparing missions and observatories, but it will provide some measure of the importance that the community attaches to the observations and has the advantage that the criteria for inclusion are narrowly controlled, while self-citation is irrelevant in this context.

### 2.4. Archival Usage

Statistics on "archival research papers" have been used as metrics for the effectiveness and impact of observatory archives. However, the concept of an archival paper is problematic, since it is hard to define and the papers even harder to identify unambiguously. For instance, a publication that has the Principal Investigator (PI) as its first author, published within three years of observation, is clearly not an archival paper. But published six years after observation, it would be. Or, if neither PI nor Co-Investigators (CoI) are on the author list, it would be, even after two years. And the same, probably, were true if only one CoI were the last among many authors. This makes it hard to identify archive papers unambiguously and, if one were to try to do it rigorously, one would seriously under-count the papers. To complicate the matter further, there are many papers that combine new observations with data from the archive. The best one can do is state that papers that present observations after four years or that re-present observations have *archival content*.

## 3. Data Used

The information in this paper reflects the state of the Chandra bibliographic database as of 10 August 2011. Where annual publication rates are used, the cut-off date is 31 December 2010. The actual queries used in this study are fairly complicated, employing joins over various databases, but the basic information is available from the Chandra statistics pages[5], bibliographic database interface[6], archive search and retrieval interface[7], and footprint service[8].

We use two types of articles in this paper: those that present observations and those that cite results; both types are restricted to refereed journal articles. For details on the selection criteria, see Section 2.1.

The rules that follow are, unfortunately, complicated, as the structure of Chandra's observing catalog was not designed for the convenience of the bibliographic database. We designed these rules to achieve an acceptable level of consistency in our data.

The publications that are used in this paper presenting Chandra observations are required to be linked to specific observation identifiers ("*ObsIds*"), representing a single contiguous exposure. Calibration observations are excluded. Annual exposure time totals are counted by *ObsId*. Where the term "observation" is used it will refer to Sequence Numbers ("*SeqNum*") which are "logical observations" consisting of one or more *ObsIds*. A link is established between a *SeqNum* and a paper when any *ObsId* in the *SeqNum* is linked to that paper. When observation and publication dates are compared, the last of the first data release dates (to the PI) of the *ObsIds* in the *SeqNum* is compared with the official publication date of the article.

We are aware that there is no satisfactory unambiguous definition of an "observation". We have chosen *SeqNum* as our counting unit since it, at least, combines the *ObsIds* of long observations that were split for scheduling reasons, rather than counting those multiple times. However, we realize that this still does not count monitoring sequences, grids, and groups (higher order observation aggregates in the Chandra mission) as single observations. On the other hand, it is not clear whether those aggregates are always treated as single observations in the literature – especially monitoring sequences. In the absence of an obviously better choice, we feel that *SeqNum* is the best one can do – or at least, a reasonable choice.

---

[5] http://cxc.cfa.harvard.edu/cda/bibstats/bibstats.html
[6] http://cxc.harvard.edu/cgi-gen/cda/bibliography
[7] http://cda.harvard.edu/chaser
[8] http://cxc.harvard.edu/cda/footprint/cdaview.html

We realize that the collection of papers presenting observations, thus defined, is a subset: there are papers presenting observations for which we have not been able to ascertain the exact observations used and, therefore, have not been able to establish any links. We continue to work on these cases, so all results reported here are subject to revision.

For the papers citing Chandra results our selection criteria are enumerated in the previous section. Although this category over-counts by including the papers that belong in the observation presenting group, but for which we have not been able to establish any links, the total may still be an undercount: we do not include here papers that cite results, but also present new observations; those papers are only included in the observation presenting category.

We have extracted five types of statistics:

- The total amount of exposure time, excluding engineering and calibration observations, obtained in each calendar year
- The total amount of exposure time represented by the publications dated in each calendar year; we extracted the total published (i.e., the sum of the exposure time represented in each paper, which may include multiple appearances of the same observation in different papers) and the total of unique exposure time (i.e., the sum of the exposure time of all *ObsIds* represented in publications that year)
- The fraction of exposure time in each calendar year that has been published – once or multiple times; and the fraction published by observation type: GO (Guest Observer), GTO (Guaranteed Time Observation), TOO (Target Of Opportunity), DDT (Director's Discretionary Time)
- The amount of time that has elapsed between the release of an observation to the PI and the first publication that presents it; and the amount of time elapsed for subsequent publications
- The number of refereed journal articles that cite Chandra results

We present these results in the next section, grouped in four themes.

## 4. Results

On the basis of an analysis of the body of statistical information we conclude that there are three areas where one can derive a definitive and meaningful statement that is quantitative and allows comparison between missions and observatories:

- The time it takes for the observations to be published
- The amount of observational data that remains unpublished (or: the percentage that gets published)
- The degree to which observational data are reused in subsequent publications

A fourth area, citing Chandra results, is informative, but lends itself less well to definitive statements.

### 4.1. Speed of Publication

Fig. 1 presents a histogram of the number of observations as a function of the amount of time elapsed between the release of the data to the PI and the first publication to present the data. The bin size is three months. It is immediately clear that the peak of the distribution is around 2 years; the median is 2.36 years. This is consistent with the almost linear increase of the published percentage of exposure time over a period of four years, from 20% to 80%, with the half-way point reached after two years (this is confirmed in subsequent figures; see, e.g., the period 2006–2010 in Fig. 5). The obvious conclusion is that it takes, on average, two years to analyze an observation and publish the result in a refereed journal. During the first few years of the mission this period was, understandably, shorter: about 18 months.

Fig. 2 presents the cumulative fraction of publication as a function of time after observation: the normalized integral over Fig. 1. This will be of interest in the next section.

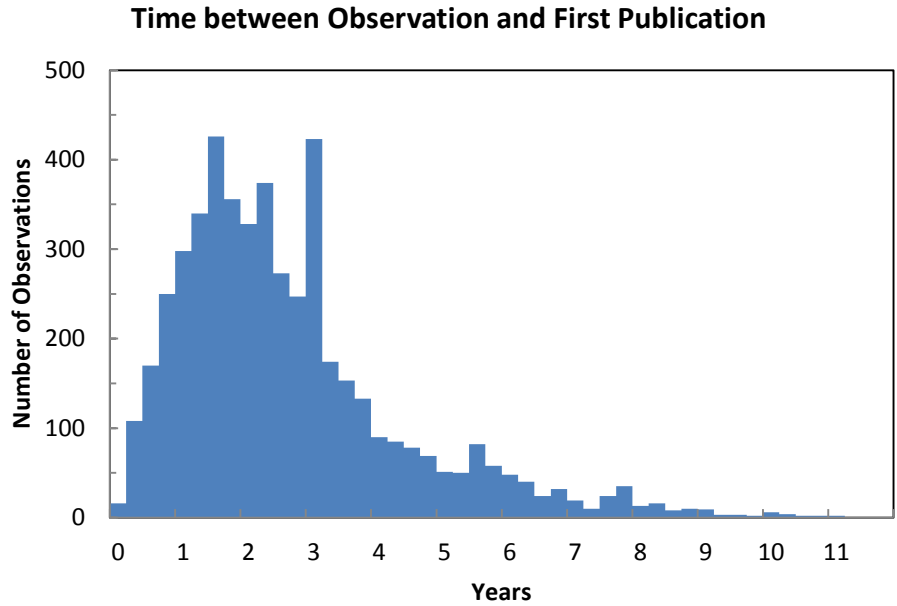

Fig. 1. Histogram of the number of observations as a function of time elapsed between the release of the data and their first publication in a refereed journal. The bin size is three months.

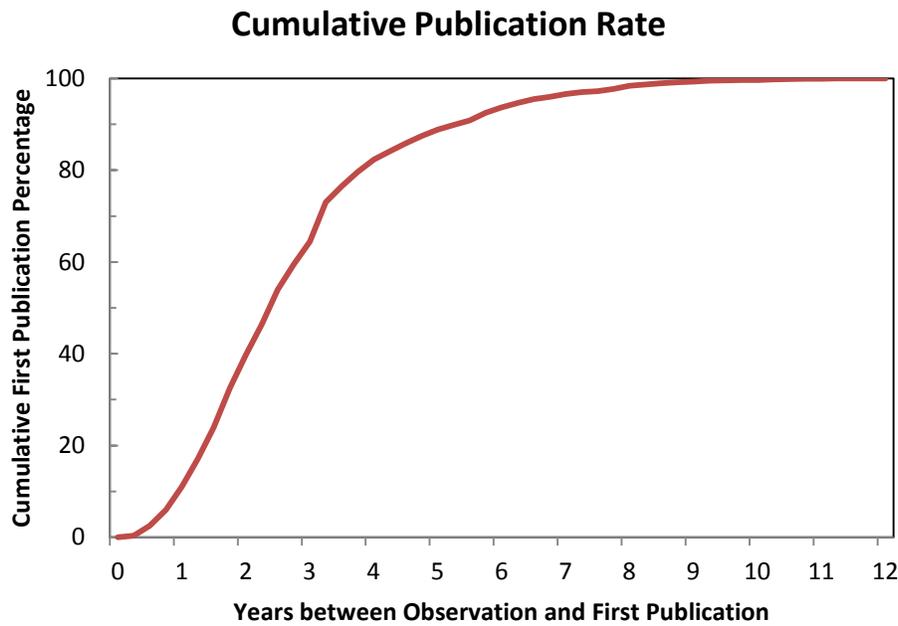

Fig. 2. The cumulative fraction of publication as a function of time lag between observation and first publication: the normalized integral over Fig. 1.

The next question is what happens with any subsequent publications. Fig. 3 provides a histogram similar to the one in Fig. 1 for these papers. It is readily understood if split into three components. The rise on the left hand side represents the regime where the first publications take place; naturally, one would expect a rising period that takes two to four years, based on the distribution in Fig. 1. Similarly, one expects a drop-off on the right hand side that takes about four years, reflecting the similar drop-off in the percentage of published exposure time. In between the two, there is a respectable plateau; it will be interesting to see whether this plateau widens and remains flat as the mission ages. However, the plot does show convincingly that there is a healthy reuse of older data from the archive and that the early observations have not lost their scientific relevance. This will be discussed further in the section on archival usage.

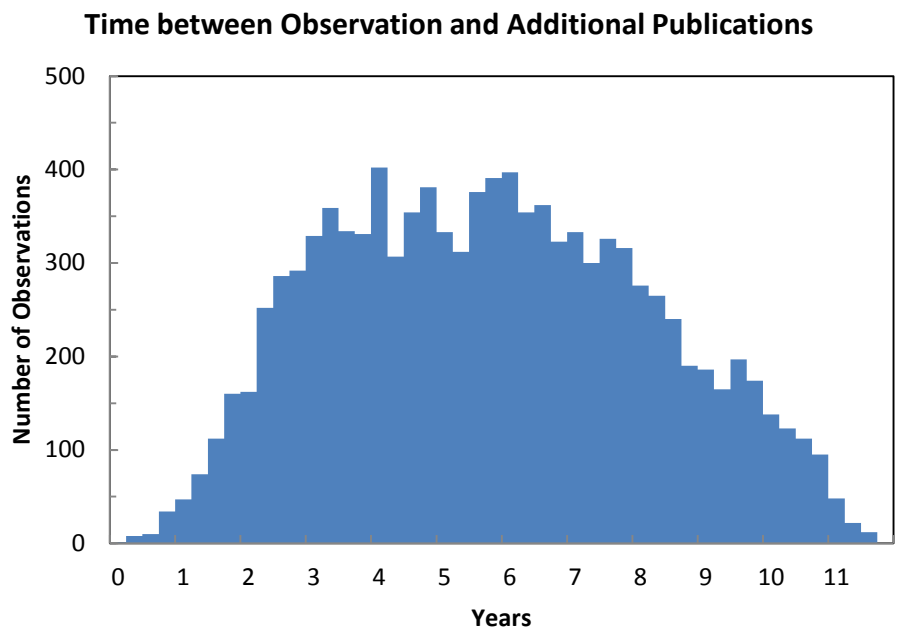

Fig. 3. Histogram of the number of observations as a function of time elapsed between the release of the data and all subsequent publications in a refereed journal. The bin size is three months.

### 4.2. Percentage of Data Published

Fig. 4 presents the percentage of unpublished exposure time, by proposal cycle, for four types of observations (GO, GTO, TOO, and DDT) as well as the total for these four categories. For GO and GTO we only show proposal cycles 1 through 7; after that cycle we run into the general publication delay (see next section).

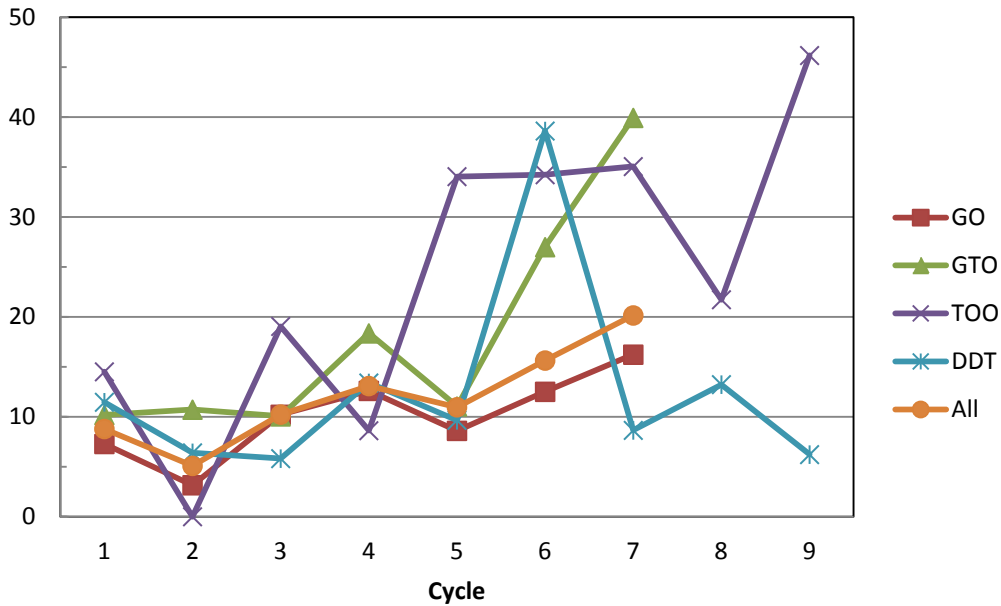

Fig. 4. Percentage unpublished observing time by proposal cycle, for four types of observations and their total.

It is worth noting that the slope of the function for all types of observations is similar to the slope in Fig. 2 for the out-years – as well as the mirror image of the outline in Fig. 5. One should keep in mind, though, that we are not comparing the same quantity: Fig. 2 refers to the fraction of the number of papers, while Fig. 4 refers to the fraction of unpublished exposure time. Nevertheless, it appears quite plausible to assume that the percentage of unpublished exposure time will gradually flatten out and approach the 10% level for all cycles.

Since TOO and DDT observations, by their nature, are published faster, we provide the percentage for these categories for two more cycles. However, the DDT percentage for cycle 9 should be regarded with caution, since a large portion of the time was dedicated to Deep Field observations which skewed the unpublished percentage downward. Because of the steep rise after cycle 4, we did compare the unpublished fraction of TOO triggered observations and TOO follow-ups for cycles 4 through 7 in various publication forms, but nothing stood out:

- The vast majority of the data are published in refereed journals; circulars and telegrams do not have a significant impact on the statistics
- Follow-up observations are doing better than the triggered TOO observations; this is not surprising, since the existence of a follow-up observation usually means that something was detected; however, even follow-up observations do not reach the 90% publication level in cycles 5 and 6, although they do again in cycle 7

It may just be that after the first few years of the mission users were less inclined to consider negative results of TOO observations significant enough to publish them.

We make the following observations:

- The fraction of unpublished exposure time will probably reach a value around 10%, at least for the first seven proposal cycles
- GTO observations have a higher unpublished percentage than GO observations for cycles 6 and 7
- The percentage of unpublished TOO data increased sharply after the first four proposal cycles
- With the exception of cycle 6, DDT observations were published at the same rate or better than the average

### 4.3. Archival Usage

One of the factors that is a measure for a mission's impact is the continued relevance and popularity of its data archive. In a previous section we concluded from the distribution in Fig. 3 that the Chandra mission is healthy in that respect. As a matter of fact, the majority of the refereed papers presenting Chandra data are publishing a re-analysis of data for which results were already published previously. This is even clearer in Fig. 5 which displays for each calendar year the percentage of exposure time that remains unpublished, is currently published once, twice, thrice, or more than three times.

Fig. 5 led us to a related, but different, statistic. Fig. 6 provides the percentage of exposure time published one, twice, thrice, and more often as a function of the data's age, in steps of one year. It looks like the mirror image of Fig. 5, but it is different in a subtle but significant way. For instance, Fig. 5 shows the percentage of exposure time from 2010 that was published after one year. Fig. 6, on the other hand, shows what percentage of exposure time was published after it had been available for one year; and that includes not just 2010, but covers all previously released data.

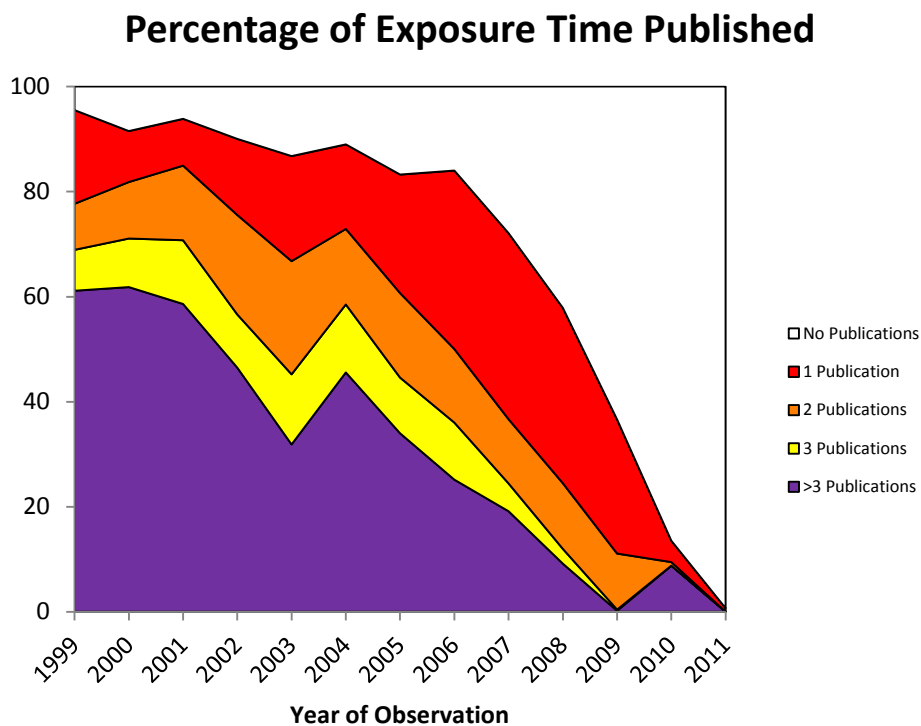

Fig. 5. Percentage of annual exposure time remaining unpublished, published once, twice, thrice, and more than three times. The purple bump in 2010 represents the Deep Field observations made during that year.

What these figures make clear is that an obvious majority of the publications presenting Chandra observations includes archival material. That is not to say that such publications should be classified as "archival research papers," but classifying them as papers "with archival content" is appropriate.

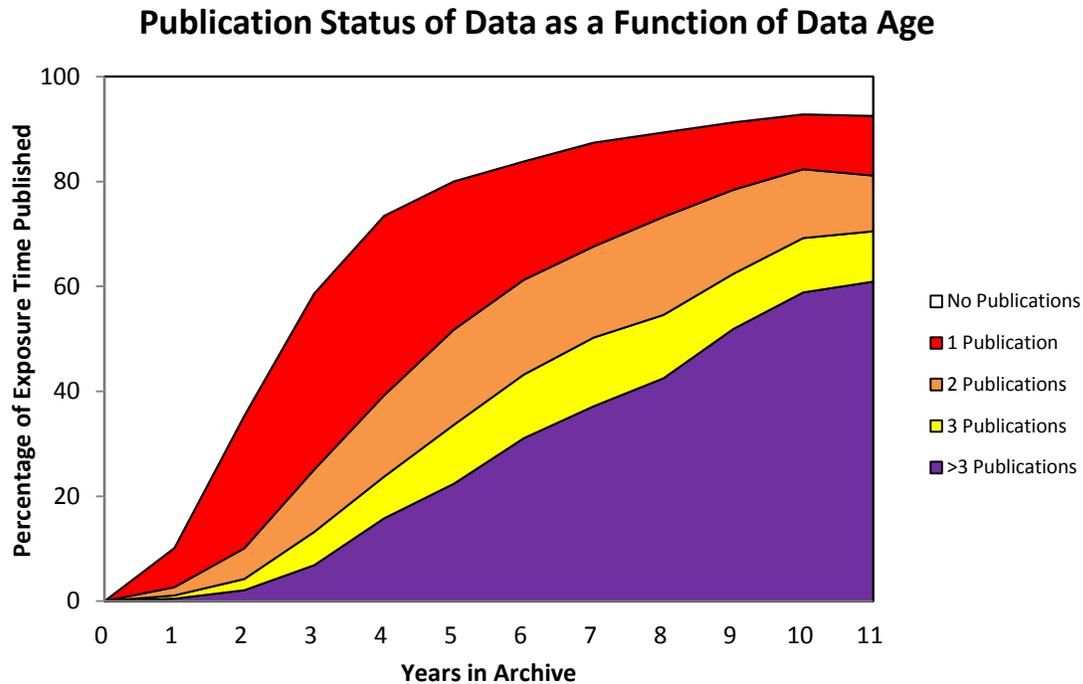

Fig. 6. Percentage of exposure time remaining unpublished, published once, twice, thrice, and more than three times, as a function of the age of the data, in annual increments.

It is still possible, though, to provide quantitative measures of archival relevance – quantitative measures that probably even allow comparisons across missions and observatories. We suggest three kinds of parameters here.

1. The median time between data becoming available and publication (excluding first publication) in Fig. 3 is 5.77 years. This is close to half the age of the mission. The caveat is, though, that this value may change over time. If one does want to make inter-observatory comparisons, one should track this value as a function of the observatory's age.
2. Turning to Figs. 5 and 6, one might devise a measure of the percentage of exposure time that has been published more than once, or maybe more than twice. But one would have to restrict that to data that are at least six or seven years old to get into a stable regime.
3. Finally, there is the ratio of the amount of exposure time published during a specific period of time and the amount of exposure time available. Keeping in mind the typical time lag between observation and publication, as evidenced in Figs. 1, 5, and 6, we shall assume that, except for the early years of the mission, it typically takes two years from the start of an investigation until the publication of a refereed paper. The metric we use in Fig. 7 is the total amount of exposure time published in each calendar year as a percentage of the cumulative amount of observing time available at the start of the preceding calendar year; i.e., the percentage for 2008 would use the amount of exposure time published in 2008, as a fraction of the exposure time available on 1 January 2007. Since publication, understandably, was faster in the early years of the mission, we used the amount available at the *end* of the preceding year for 2003 and earlier. There are two choices for this metric: the amount of exposure time as the sum of the exposure time presented in each paper ("All Data"); or the sum of the exposure time of all unique observations published in these papers ("Unique Data"). The former tracks the total scientific effort and hovers between 60% and 70%. The latter is an indication of what percentage of the available data is of special interest at any given time; it is remarkably constant at 40%. This is consistent with the tentative conclusion by Apai et al. (2010) that archival publications appear to be proportional to the total content of the archive.

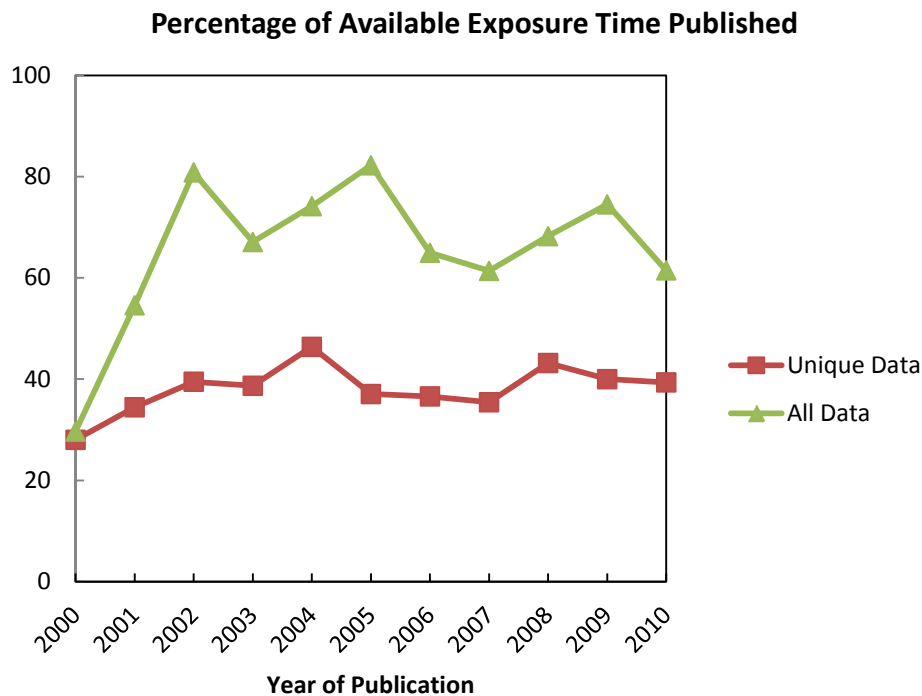

Fig. 7. Percentage of exposure time published in refereed papers during a single calendar year, relative to the amount of exposure time available at the end (pre-2004) or the start (post-2003) of the preceding year. See text for details.

### 4.4. Citation of Results

As argued in the section on metrics, if one feels the need for an impact metric based on citations, a better choice is to count the number of refereed journal articles that cite substantive results, since it allows more precise control over the criteria defining the metric.

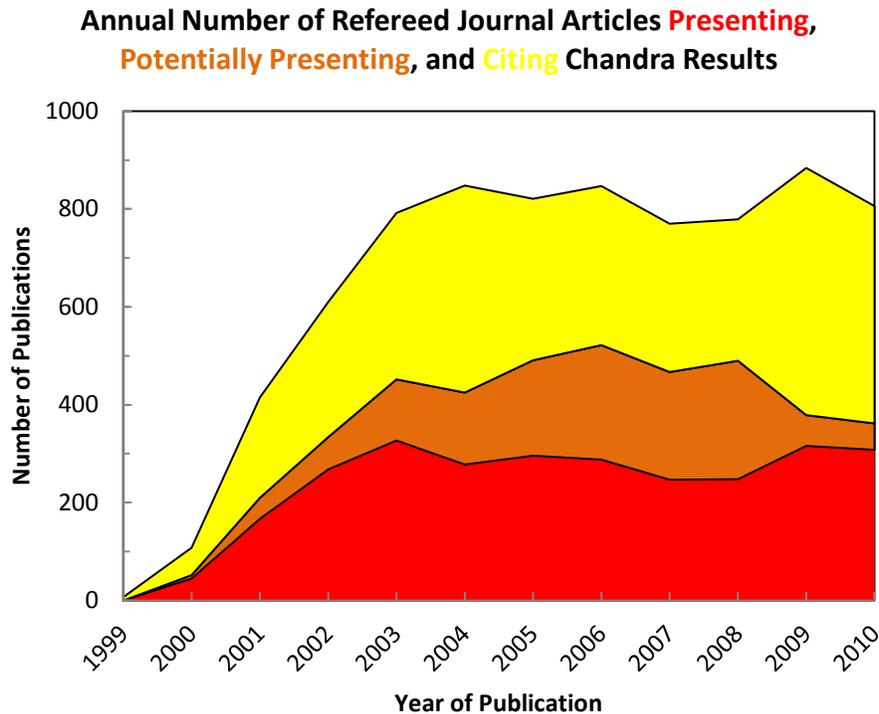

Fig. 8. Stacked plot of the annual number of refereed journal articles that present Chandra observations and have links to the datasets (red); articles that probably present observations, but for which we have not (yet) been able to establish data links (orange); and articles that cite Chandra observations or results derived from them (yellow).

Fig. 8 presents a stacked plot of the annual numbers for three types of refereed journal articles: articles that cite Chandra results; articles that cite results and may be presenting observations, but for which we do not (yet) have definite data links; and papers that present specific Chandra observations. This last set represents the articles used in the other statistics in this paper; as explained, their actual number is far less informative than the amount of exposure time that they actually present and one should not attach more weight to these numbers than they deserve.

# 5. Conclusions

## 5.1. The State of Chandra in Publications

We summarize the results of this study:

- The median time it takes for a Chandra observation to get published is 2.36 years. This is consistent with the time it takes for a year's worth of observations to reach the flat part of the fraction published curve (4 years) and its half-way point (2 years).
- The median time for subsequent publications is 5.77 years. This is half the life of the mission, but time will tell whether that is significant.
- Roughly 10% of the exposure time remains unpublished in the long term. The fraction of unpublished TOO data increased sharply after proposal cycle 4.
- A significant majority of Chandra publications includes archival content; as much as 60% of all exposure time may eventually be presented in publications more than twice.
- During any given year, about 40% of all available exposure time is extracted to be analyzed and published or re-analyzed and re-published within two years. The total annual publication output of the mission, using that same metric, is between 60% and 70% of the cumulative observing time. For 2010 these metrics amount to almost four times and six times, respectively, the annual exposure time budget of the observatory.
- The results of the mission are cited in at least as many refereed papers as those that present the observations.

The one caveat on the numbers here presented is that we continue to work on establishing data links to papers for which we have, so far, been unsuccessful. Consequently, the exact numbers are subject to revision.

We conclude that the publication activity associated with the Chandra mission is healthy and vigorous. Archival data constitute a substantial part of the contents of these publications. At this time data from the entire life of the mission remain relevant.

## 5.2. Usable Metrics

On the basis of this analysis we suggest that the following metrics may be useful to other observatories and possibly (though that does not necessarily follow) can be used to make comparisons between observatories. Key to these metrics is the linking between papers and observational datasets in bibliographic databases.

### 5.2.1. Percentage of Data Published

- The percentage of exposure time that is presented in refereed publications; this should be based on observations that are older than about three or four times the median time it takes to publish observations (cf. Figs. 1, 2, 6)

### 5.2.2. Speed of Publication

- Median time between the date that the data of an observation become available and the first publication to present that observation (cf. Figs. 1 and 2)
- The time it takes to reach the stable publication level defined as percentage published (see above; cf. Figs. 2, 5, 6)
- The time it takes to reach the half-way point to the stable publication level (cf. Figs. 2, 5, 6)
- Alternatively, one might use the time it takes to reach 20%, 40%, 60%, 80% of the total exposure time (cf. Fig. 6)

### 5.2.3. Archival Usage

- Median time between the date that the data of an observation become available and the dates of subsequent publications, after the first; this median could be expressed as a fraction of the observatory's age and should be tracked as a function of that age (cf. Fig. 3)
- Percentage of available exposure time that is published in a calendar year, both total and unique, with an appropriate lag based on publication speed (cf. Fig. 7)
- Percentage of exposure time that is ten years old and published more than twice (cf. Fig. 6); an age of ten years is a fairly arbitrary estimate and it is possible that it should be expressed as a multiple of the median publication delay from Fig. 1; this metric might serve as a measure of the observatory's impact.

### 5.2.4. Citation of Results

- The number of refereed journal articles that cite results derived from the observatory's observations that are significant in the scientific case brought by such a paper (cf. Fig. 8). It may need to be normalized by the number of articles that present observations. But it is not clear this metric is suitable for cross-observatory comparison.

## Acknowledgments

This research was supported by NASA contact NAS8-03060 (CXC) and relies heavily on the services provided by the Astrophysics Data System (ADS). We gratefully acknowledge the work by all past and current members of the Chandra Data Archive Operations Team who have helped build and maintain our bibliographic database. We thank the bibliographic curators at ESO, HEASARC, HST, NRAO, and Spitzer for providing journal usage data. And finally, we are indebted to Drs. H. Tananbaum and P. Green, and to the anonymous referee, for providing helpful suggestions.